Phantom chain simulations for fracture of star polymer networks with various strand densities

*Yuichi Masubuchi, Takato Ishida, Yusuke Koide, and Takashi Uneyama

Department of Materials Physics, Nagoya University, Nagoya 4649603, Japan

*To whom correspondence should be addressed mas@mp.pse.nagoya-u.ac.jp

Ver Jul 14, 2024

**Abstract**

Despite many attempts, the relation between fracture and structure of polymer networks is yet to be clarified. For this problem, a recent study for phantom chain simulations [Macromolecules, 56, 9359 (2023)] has demonstrated that the fracture characteristics obtained for polymer networks with various node functionalities and conversion ratios lie on master curves if they are plotted against cycle rank. In this study, we extended the simulation to the effect of prepolymer concentration on the relationships between cycle rank and fracture characteristics within the concentration range of $1 \lesssim c/c^* \lesssim 8$, concerning the overlapping concentration $c^*$. We created networks from sols of star-branched phantom bead-spring chains via end-linking reaction between different chains through Brownian dynamics simulations with varying the number of branching arms $f$ from 1 to 8, and the conversion ratio $\varphi_c$ from 0.6 to 0.95. For the resultant networks, cycle rank $\xi$ was consistent with the mean-field theory. The networks were uniaxially stretched with energy minimization until break to obtain modulus $G$, strain at break $\varepsilon_b$, stress at break $\sigma_b$, and work for fracture $W_b$. With the branch point density $v_{br}$, $G/v_{br}$, $\varepsilon_b$, $\sigma_b/v_{br}$, and $W_b/v_{br}$ of the data for various $f$ and $\varphi_c$ draw master curves if plotted against $\xi$. The master curves depend on $c$; as $c$ increases, all the mechanical characteristics monotonically increase. If we plot $\sigma_b/v_{br}$ and $W_b/v_{br}$ against $G/v_{br}$, the data for various $f$ and $\varphi_c$ lie on master curves but depending on $c$. Consequently, the fracture characteristics are not solely described by modulus for the examined energy-minimized phantom chain networks.

**Introduction**

More needs to be clarified about which structural characteristics dominate the fracture of polymer networks[1]. A few theoretical attempts have been reported focusing on the effect of loops and crack propagation. Barney et al.[2] have extended the real elastic network theory (RENT)[3] to describe the fracture energy, which decreases with increasing the number of loops. Their approach relies on the Lake-Thomas theory[4], in which fracture energy is calculated from the work for crack propagation in tearing tests, and the change in fracture energy essentially comes from the modulus. They demonstrated that the theoretical prediction is consistent with the results of coarse-grained molecular simulations, in which they varied the conversion rate and the strand molecular weight. Lin and Zhao[5] reported a different theoretical attempt in which they calculated fracture energy considering crack tip propagation. According to this theory, fracture energy increases with secondary loops (cyclic loops) increase, contradicting Barney et al. [2] There have been reported several experimental results that cannot be described by these theories. For instance, Akagi et al.[6] observed the fracture of tetra-PEG gels created with various conversion ratios and prepolymer concentrations. Because their networks are made from mixtures of star polymers where each molecule reacts only with the other, primary and higher odd-ordered loops are omitted. Although they observed fracture under uniaxial stretch without an initial notch, the fracture energy and modulus correlate for gels with various conversion ratios, similar to the Lake-Thomas theory. However, the prepolymer concentration dependence is not solely described by modulus. The other example reported by Fujiyabu et al.[7] is that gels made from 3-arm star prepolymers exhibit better fracture properties than those for 4-arm analogs, even if the modulus is the same. They explained this difference by stretch-induced crystallization of the 3-arm case. However, Masubuchi et al.[8] performed coarse-grained simulations to demonstrate that the superiority of 3-arm gels is seen even without crystallization.

An interesting clue for network fracture is cycle rank. Masubuchi et al.[9] performed coarse-grained simulations for elastic networks prepared from star branch prepolymers to obtain strain and stress at break and work for fracture under uniaxial elongations. They found that for the networks with various branch functionalities and conversion ratios with monodisperse arm length, some fracture characteristics follow master curves if plotted against the cycle rank. Masubuchi[10] also showed that fracture characteristics for the networks created from mixtures of star polymers with different functionalities lie on the same master curves. In the subsequent work, Masubuchi[11] further examined end-linking networks composed of linear prepolymers and multi-functional linkers to report that the relationship between fracture characteristics and cycle rank is essentially the same as that for star polymer networks despite including loops.

This study used phantom chain simulations to explore the relationship between fracture properties and cycle rank for cases with different prepolymer concentrations. Consistent with earlier studies[1,8,12–14], the simulation results showed that modulus and fracture energy increase with increasing prepolymer concentration. The master curves concerning the relation between fracture properties and cycle rank hold for various prepolymer functionalities and conversion ratios, but they depend on prepolymer concentration. The fracture properties depend on prepolymer concentration differently from that of modulus. Details are shown below.

**Model and Simulations**

The employed model and simulation scheme is shared with the previous studies[8–11,15] except for the prepolymer density. Examined networks were created from equilibrated sols of phantom star chains with various concentrations through the Brownian simulations with end-linking reactions. The obtained networks were energy-minimized and stretched until the break. During the elongation, the evolution of stress was recorded as a function of strain, and from the stress-strain relationship thus obtained, the fracture characteristics were extracted.

The prepolymers are represented by bead-spring chains, for which $f$-arms are connected to the central bead, and the bead number of each arm is $N_a$. A sufficiently large number of prepolymers $M$ were dispersed with several bead number densities $\rho$ in a cubic simulation box with periodic boundary conditions and equilibrated via the Brownian dynamics. The equation of motion for the position of each bead $\mathbf{R}_i$ is written as follows.

$$0 = -\zeta \dot{\mathbf{R}}_i + \frac{3k_B T}{a^2} \sum_k f_{ik} \mathbf{b}_{ik} + \mathbf{F}_i \quad (1)$$

The first term on the right-hand side is the drag force, and $\zeta$ is the friction coefficient. The second term is the elastic force generated by springs. Here, $a$ is the average bond length, $\mathbf{b}_{ik}$ is the bond vector defined as $\mathbf{b}_{ik} \equiv \mathbf{R}_i - \mathbf{R}_k$ and $f_{ik}$ is the non-linear spring constant written as $f_{ik} = (1 - \mathbf{b}_{ik}^2/b_{\max}^2)^{-1}$ with the maximum stretch $b_{\max}$. This non-linear spring constant with finite extensibility avoids bond extension due to thermal fluctuations. $k_B$ is the Boltzmann constant and $T$ is temperature. The third term is Gaussian random force that obeys fluctuation-dissipation relation with the first term. Since no inter-bead interactions are considered, chain overlapping and crossing are allowed. For the employed model, units of length, energy, and time are defined from this Brownian scheme as $a$, $k_B T$, and $\tau = \zeta a^2/k_B T$. The quantities are normalized according to these units, hereafter. Eq 1 was numerically integrated by a second-order scheme[16] with the step size $\Delta t$.

After sufficient equilibration, end-linking reactions were turned on[17,18]. Following the experiments for tetra-PEG type gels[1,12], the prepolymers were binary labeled, and the reaction took place only between prepolymers having different labels. Thus, no primary and higher odd-order loops were included in the network, whereas secondary and higher even-order loops were naturally created. The reaction occurs with the probability $p$ when a pair of reactive ends come closer than the critical distance $r_c$. During the gelation, snapshots of the system with various conversion ratios $\varphi_c$ were stored.

The networks thus obtained for various combinations of $(\rho, f, \varphi_c)$ values were uniaxially stretched. The stretch was performed for energy-minimized networks, as done in earlier studies[19–22]. The total energy is written below and is consistent with the non-linear spring employed in the Brownian scheme.

$$U = -\frac{3k_B T b_{\max}^2}{2a^2} \sum_{i,k} \ln\left(1 - \frac{\mathbf{b}_{ik}^2}{b_{\max}^2}\right) \quad (2)$$

This energy was minimized with the Broyden-Fletcher-Goldfarb-Sanno method[23], in which the beads were moved within an infinitesimal distance $\Delta r$ without Brownian motion according to the potential gradient until the total energy converged to a specific value within a given allowance $\Delta u$. In this energy-minimized structure, elongated bonds were removed when the bond length became larger than a certain critical length $b_c$. The energy-minimization and stretch steps were alternatively repeated until the network percolation was eliminated.

The advantage of the employed energy-minimized scheme is that the results are not affected by the number of parameters such as the elongation rate and bond cutting criterion, as discussed previously[8]. The drawback is the lack of thermal motion, and we cannot discuss the energy dissipation according to the structural relaxation[24].

Since the primary purpose of this study is to discuss the effect of prepolymer concentration, the bead number density $\rho$ was varied to 2, 4, and 16, and the results shall be compared with those for the case of $\rho = 8$ examined in the previous study[9]. These densities correspond to $1 \lesssim c/c^* \lesssim 8$, with respect to the overlapping concentration $c^*$. The other parameters were the same as in the previous studies. $3 \leq f \leq 8$, $N_a = 5$, $M = 1600$, $\Delta t = 0.01$, $b_{max} = \sqrt{1.5}$, $b_c = \sqrt{2}$, $p = 0.1$, $r_c = 0.5$, $0.6 \leq \varphi_c \leq 0.95$, $\Delta r = 0.01$, and $\Delta u = 10^{-4}$. Eight independent simulation runs were performed for each condition, and the quantities reported below are ensemble averages unless stated.

**Results and Discussion**

Figure 1 shows snapshots of examined networks before and after energy minimization for $f = 4$ at $\varphi_c = 0.95$ with various $\rho$ values. Since the number of prepolymers $M$ is fixed at 1600, the simulation box size decreases with increasing $\rho$. Even for the largest $\rho$ with the smallest box case, the box dimension is large enough compared to prepolymers; the box size is ca. $12.8^3$ for $f = 4$ whereas the average end-to-end distance is $\sqrt{10}$. Note that $M=1600$ is sufficient for fracture simulations as demonstrated previously[8]. Nevertheless, the effect of system size shall be discussed later.

Concerning the network structure, when the polymer concentration is low, density inhomogeneity due to kinetic arrest is observed, as experimentally reported earlier[25]. Energy minimization enhances this inhomogeneity. Nevertheless, the network structures thus obtained are similar to those reported in earlier studies[19–22].

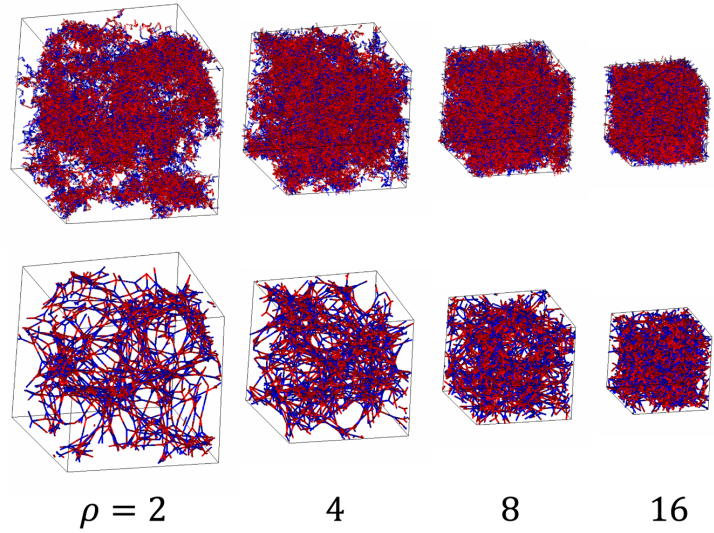

$\rho = 2$      4      8      16

**Figure 1** Typical snapshots of examined networks after gelation (top row) and after energy minimization (bottom row) for $f = 4$ at $\varphi_c = 0.95$ with $\rho = 2, 4, 8$, and 16, from left to right. This range of $\rho$ corresponds to $1 \lesssim c/c^* \lesssim 8$. Blue and red indicate different chemistries between which the end-linking reaction occurs.

Figure 2 shows cycle rank $\xi$ as a function of $\varphi_c$. $\xi$ does not depend on $\rho$, and the $\varphi_c$-dependence is entirely consistent with the mean field theory[26,27], as previously reported[9] for the case of $\rho = 8$. This coincidence demonstrates that each reaction occurs independently despite structural inhomogeneity in Fig 1. However, the correspondence of $\xi$ among the networks with different $\rho$ does not mean the maturity of the effective network. Namely, inert fractions like dangling domains are not discriminated in calculating $\xi$, whereas unconnected portions to the percolated networks are excluded.

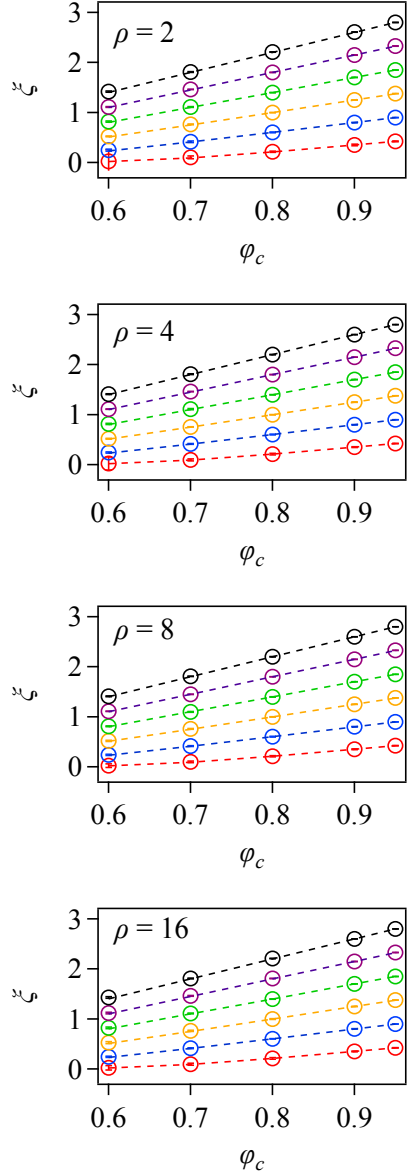

**Figure 2** Cycle rank per branch point $\xi$ plotted against $\varphi_c$ for $\rho = 2, 4, 8$, and $16$ from top to bottom with $f = 3$ (red), $4$ (blue), $5$ (orange), $6$ (green), $7$ (violet), and $8$ (black). Dotted curves indicate the mean-field calculation. Error bars (within symbols) show the standard deviation of 8 different simulation runs. Note that the results for $\rho = 8$ were reported previously.

Figure 3 shows examples of the development of true stress $\sigma$ during elongation against true strain $\varepsilon$. For this specific case, $f = 4$, $\varphi_c = 0.95$, and $\rho = 2, 4, 8$, and $16$. The modulus $G$ increases with increasing $\rho$ since the density of effective strands increases. During elongation, stress fluctuates, reflecting the breakage of some strands. Owing to the energy-minimization scheme, stress immediately goes down to zero when network percolation is eliminated. Conversely, energy dissipation due to network relaxation[24] is not considered. Nevertheless, in addition to modulus ($G$), strain and stress at break ($\varepsilon_b$ and $\sigma_b$) and work for fracture ($W_b$) were obtained from each stress-strain curve. These values were averaged among 8 independent simulations for various $\rho$, $f$, and $\varphi_c$, and the data are shown below.

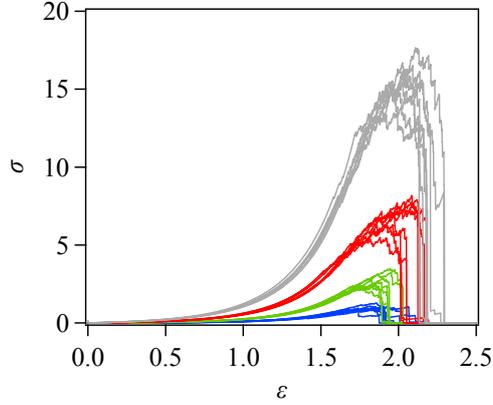

**Figure 3** True stress versus true strain during elongation for $f = 4$ and $\varphi_c = 0.95$. The results for $\rho = 2, 4, 8,$ and $16$ are shown by blue, green, red, and gray curves, respectively. Each curve corresponds to a single simulation run.

Figure 4 shows the modulus $G$ plotted against $\rho$ for $\varphi_c = 0.6$ (top) and 0.95 (bottom) with various $f$. The modulus was obtained as the value of $\sigma/(\lambda^2 - \lambda^{-1})$ with the stretch $\lambda$ at $\lambda^{-1} = 0.75$, as in the previous study[9]. As mentioned in Fig 3, $G$ increases with increasing $\rho$. However, $G$ is not proportional to $\rho$ for low $\rho$. See the deviation from the dotted lines that indicate the slope of unity. This non-linearity of $G$ against $\rho$ reflects that some fractions of network nodes do not effectively sustain stress.

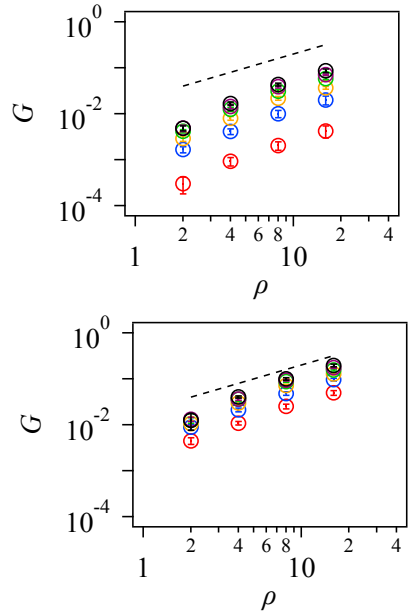

**Figure 4** Modulus $G$ plotted against $\rho$ for $f = 3$ (red), 4 (blue), 5 (orange), 6 (green), 7 (violet), and 8 (black) at $\varphi_c = 0.6$ (top) and 0.95 (bottom). Broken lines indicate the slope of unity. Error bars

correspond to the standard deviations for 8 different simulation runs.

Figure 5 shows $G$ normalized by the number density of branch points $v_{br}$ plotted against cycle rank $\xi$ for various $\rho$. (Here, $v_{br}$ is calculated for the entire system, and unreacted prepolymers and dangling ones are included.) Consistent with the phantom network theory[28], the data for various $f$ and $\varphi_c$ lie on a master curve for each $\rho$, as reported previously for $\rho = 8$. The weak non-linearity in the small $\xi$ region is due to strand-extending prepolymers, which have only two reacted arms and extend the strand length between mechanically effective network nodes[9]. Concerning the effects of $\rho$, if the networks are perfectly connected, as assumed in the phantom network theory[28], this normalized plot should be independent of $\rho$. Indeed, the results for $\rho = 8$ (red) and 16 (green) are almost the same. However, for lower $\rho$ values, the modulus becomes smaller, reflecting ineffective fractions of the network.

One may argue that small modulus $G$ for low $\rho$ is inconsistent with cycle rank $\xi$ shown in Fig 2, where $\xi$ is entirely consistent with mean-field theory irrespective of $\rho$. This difference between $G$ and $\xi$ for $\rho$-dependence implies that these two parameters characterize different aspects of networks. Namely, $G$ reflects a part of networks that are involved in force paths, whereas $\xi$ is for connectivity on average, irrespective of force propagation.

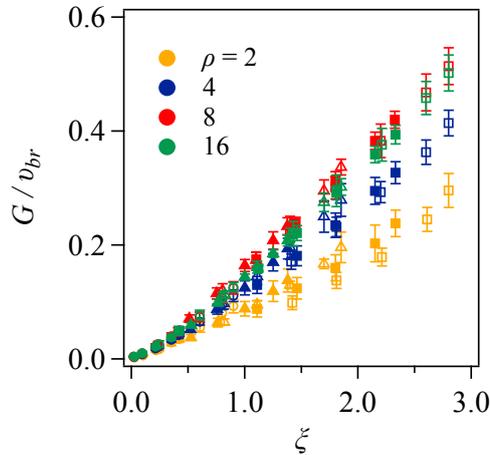

**Figure 5** Modulus $G$ normalized by the number density of branch points $v_{br}$ as a function of cycle rank $\xi$ for various $\rho$. Symbols indicate $f = 3$ (filled circle), 4 (unfilled circle), 5 (filled triangle), 6 (unfilled triangle), 7 (filled square), and 8 (unfilled square). Error bars correspond to the standard deviations for 8 different simulation runs.

Figure 6 shows the fracture characteristics plotted against $\rho$ for $\varphi_c = 0.6$ (left column) and 0.95 (right column). Concerning $\varepsilon_b$ shown in panels (a1) and (a2), Akagi et al.[6] reported for tetra-PEG gels ($f = 4$) with $\varphi_c \sim 0.9$ that stretch at break $\lambda_b (= \exp(\varepsilon_b))$ exhibits a power-law dependence on the prepolymer concentration $\varphi$ with the power-law exponent of 1/3 ($\lambda_b \sim \varphi^{1/3}$). Since the bead number density $\rho$ is proportional to $\varphi$, their result corresponds to $\varepsilon_b = A + (1/3)\ln\rho$ with a constant $A$. This relation is shown in Panels (a1) and (a2) by broken curves with $A = 1.2$. (Note that this choice of A is ) The simulation results indicated by symbols are inconsistent with this relation. For the case of $\varphi_c = 0.6$ (panel a1), $\varepsilon_b$ is almost constant within the examined range of $\rho$ for $f > 3$. For $f = 3$ (red symbol), $\varepsilon_b$ significantly decreases with increasing $\rho$. This behavior is probably due to the difference in $\varphi_c$; no experimental report can be found for networks with such a low conversion rate.

The discrepancy for the $\rho$-dependence of $\varepsilon_b$ from the experimental reports for tetra-PEG gels is partly explained below. In the experiment, polymers are dispersed in a good solvent, and gel networks swell due to osmotic pressure. Such a swelling attains the development of a mechanically effective network with relatively homogeneous structures, even under low polymer concentrations. In contrast, our simulations are for ideal chains, and the effect of osmotic force is neglected. Thus, when the prepolymer concentration is low, structural inhomogeneity is enhanced as the reaction proceeds, as seen in Fig 1. The mechanical behaviors reflect this structural difference, at least partly. Note that even for tetra-PEG gels, structural inhomogeneity is observed when prepolymer concentration is extremely low[25].

Figure 6 panels b and c exhibit $\sigma_b$ and $W_b$ plotted against $\rho$ for various $f$, demonstrating that these fracture characteristics increase with increasing $\rho$. Even though the simulation setting is different from tearing tests, if we assume the Lake-Thomas theory, $W_b$ is expected proportional to $\rho$. Broken lines in Panels (c1) and (c2) show this behavior, with which the simulation results shown by symbols are qualitatively consistent. However, some discepancies are also observed. For instance, the dependence of $W_b$ is somewhat more i tense than the broken line when $\rho$ is low, and $f$ is large. $W_b$ for the case with $f = 3$ and $\varphi_c = 0.6$ (red circle in panel c1) saturates in high $\rho$ regime.

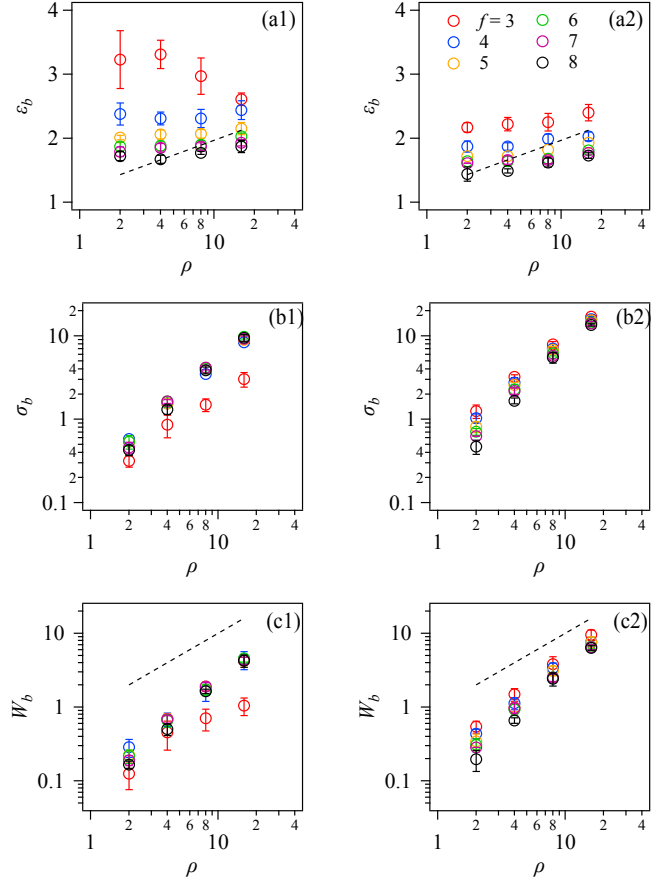

**Figure 6** Strain at break $\varepsilon_b$ (panels a), stress at break $\sigma_b$ (panels b), and work for fracture $W_b$ (panels c) plotted against the bead number density $\rho$ at $\varphi_c = 0.6$ (left column) and 0.95 (right column) for $f = 3$ (red), 4 (blue), 5 (orange), 6 (green), 7 (violet), and 8 (black). Error bars correspond to the standard deviations for 8 different simulation runs. Broken curves in panel a show $\varepsilon_b = 1.2 + (1/3)\ln\rho$. Broken lines in panel c indicate the slope of unity.

One might argue that the results depicted in Fig. 6 reflect not just the effect of density but also of system size, given that the simulation box dimensions varied, as shown in Fig. 1. For this matter, we present results with a fixed box dimension, filled with varying numbers of molecules. Figure 7 illustrates $\varepsilon_b$, $\sigma_b$, and $W_b$ for $f = 4$ at $\varphi_c = 0.6$ and 0.95, across various $\rho$, with a constant box volume in all instances. The results align with those in Fig. 6, where the number of molecules was fixed, thereby indicating that the effect of system size is insignificant.

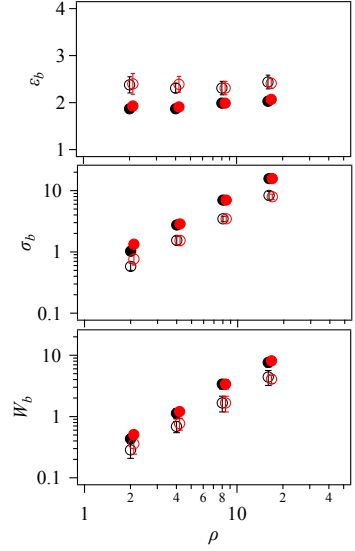

**Figure 7** Strain at break $\varepsilon_b$, stress at break $\sigma_b$, and work for fracture $W_b$ from top to bottom plotted against the bead number density $\rho$ for $f = 4$ at $\varphi_c = 0.6$ (unfilled circle) and 0.95 (filled circle) for the simulation with a constant volume (red) and with a constant prepolymer number (black). For the constant volume simulation, the volume $V$ was fixed at 4200, and the number of prepolymers $M$ was varied as 400, 800, 1600, and 3200. For the simulation with the constant prepolymer number, $M = 1600$ and $V = 16800, 8400, 4200$, and 2100. Error bars correspond to the standard deviations for 8 different simulation runs and are sometimes smaller than the symbols.

Figure 8 exhibits all the obtained fracture characteristic values for various sets of $\rho$, $f$, and $\varphi_c$ plotted against $\xi$. For each $\rho$, $\varepsilon_b$ data in panel (a) are located on a master curve that is apparently described as a power-law decay function of $\xi$; $\varepsilon_b = \varepsilon_0 \xi^{-\alpha_\varepsilon}$. As $\rho$ increases, $\varepsilon_b$ slightly increases systematically. As reported previously[9–11], $\sigma_b$ and $W_b$ for various $f$ and $\varphi_c$ also lie on master curves if these values divided by $v_{br}$ are plotted against $\xi$. These behaviors are apparently fitted by power-law functions of $\xi$; $\sigma_b/v_{br} = (\sigma_0/v_{br})\xi^{\alpha_\sigma}$ and $W_b/v_{br} = (W_0/v_{br})\xi^{\alpha_W}$. As $\rho$ increases, both $\sigma_b/v_{br}$ and $W_b/v_{br}$ increases. The master curves have been previously reported for $\rho = 8$, but they are found for the first time for the other $\rho$ values. Note that no theory has been found for the power-law functions, and the fitting is eye-guide. Note also that in some previous studies, $\sigma_b$ and $W_b$ were divided by the broken strand fraction $\varphi_{bb}$. However, the value of $\varphi_{bb}$ is hardly experimentally accessible, and $v_{br}$ was found to achieve the master curves instead of $\varphi_{bb}$ in the previous study[11].

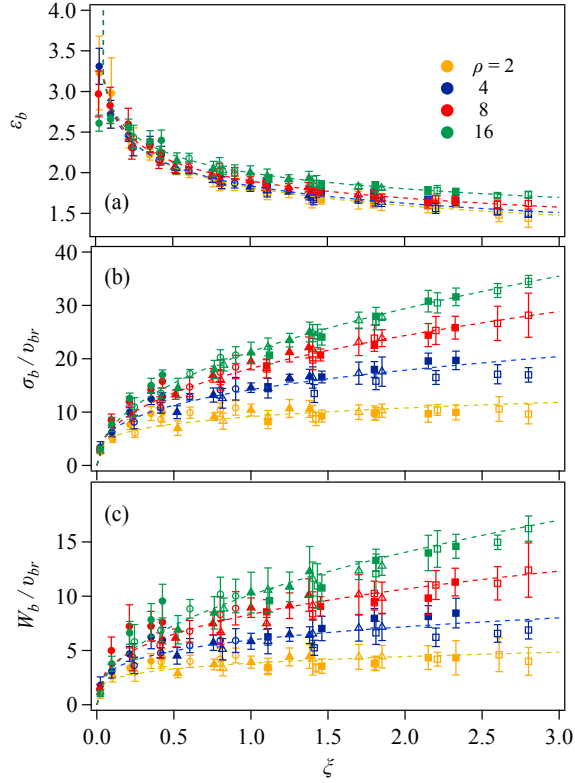

**Figure 8** Fracture characteristics $\varepsilon_b$, $\sigma_b$, and $W_b$ obtained for various $\rho$, $f$, and $\varphi_c$ values as functions of $\xi$. $\sigma_b$ and $W_b$ are normalized by the branch point density $\nu_{br}$. The $f$ values are 3 (filled circle), 4 (unfilled circle), 5 (filled triangle), 6 (unfilled triangle), 7 (filled square), and 8 (unfilled square), respectively. Error bars are standard deviations among eight different simulation runs. Broken curves show power-law functions as eye-guide; $\varepsilon_b = \varepsilon_0 \xi^{-\alpha_\varepsilon}$, $\sigma_b/\nu_{br} = (\sigma_0/\nu_{br})\xi^{\alpha_\sigma}$ and $W_b/\nu_{br} = (W_0/\nu_{br})\xi^{\alpha_W}$.

One may argue that Figures 5 and 8 suggest a correlation between the modulus and the fracture characteristics. Figure 9 examines such an argument. For each $\rho$ case, $\sigma_b/\nu_{br}$ and $W_b/\nu_{br}$ are single-valued functions of $G/\nu_{br}$. Namely, the fracture behavior for the networks with various $f$ and $\varphi_c$ is dominated by the modulus, given that $\rho$ is common. However, when $\rho$ differs, networks sharing the same $G/\nu_{br}$ value exhibit different $\sigma_b/\nu_{br}$ and $W_b/\nu_{br}$ values; these values increase with increasing $\rho$. Consequently, the fracture characteristics are not solely described by modulus. These results are consistent with the experiment by Akagi et al.[6]

The results shown in Fig 9 imply that network maturity evaluated by modulus differs from that affects fracture properties. Modulus is used to quantify the density of effective strands and nodes, and their mechanical contributions are evaluated as an averaged value. In contrast, fracture is initiated by scission of the most elongated strand and propagates to next ones. Since we conducted simulations with step-by-step energy minimization, this propagation occurs within the elongated tail of the strand length distribution. Therefore, modulus and fracture reflect different characteristics of the strand length statistics that are unnecessarily correlated. Meanwhile, we note that the energy-minimization scheme employed here may be different from that realized experimentally for gels and other polymer networks.

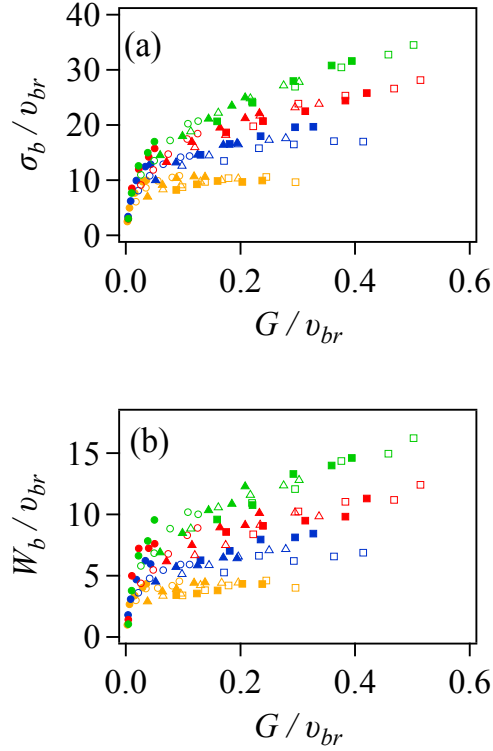

**Figure 9** Fracture characteristics $\sigma_b/v_{br}$ (a) and $W_b/v_{br}$ (b) obtained for various $\rho$, $f$, and $\varphi_c$ values plotted against $G/v_{br}$. The $f$ values are 3 (filled circle), 4 (unfilled circle), 5 (filled triangle), 6 (unfilled triangle), 7 (filled square), and 8 (unfilled square), respectively. The $\rho$ values are 2 (yellow), 4 (blue), 8 (red), and 16 (green), respectively.

**Conclusions**

We conducted phantom chain simulations to investigate the effect of prepolymer concentration on the fracture behavior of star polymer networks. We varied the star polymer functionality $f$ from 3 to 8, the conversion ratio $\varphi_c$ from 0.6 to 0.95, and the segment number density $\rho$ from 2 to 16. The resultant networks were statistically evaluated regarding the cycle rank $\xi$, which was consistent with the mean-field theory, demonstrating that the end-linking reactions occurred independently. We uniaxially stretched the networks and obtained modulus $G$ and fracture characteristics including strain at break $\varepsilon_b$, stress at break $\sigma_b$, and work for fracture $W_b$ from the stress-strain relation until the break. Consistent with experiments, these mechanical characteristics increase with increasing $\rho$. With the branch point density $v_{br}$, if we plot $G/v_{br}$, $\varepsilon_b$, $\sigma_b/v_{br}$, and $W_b/v_{br}$ against $\xi$, the data for various $f$ and $\varphi_c$ are located on master curves, as reported previously. However, different curves are realized for different $\rho$ values. We also found that if we plot $\sigma_b/v_{br}$ and $W_b/v_{br}$ against $G/v_{br}$, master curves are seen for various $f$ and $\varphi_c$, but they depend on $\rho$. This result demonstrates that fracture is not solely dominated by modulus but depends on prepolymer density.

Although the presented study gives fundamental information for network fracture, we should note that the results may change if we consider osmotic force, excluded volume interactions, and thermal fluctuations. Besides, we have not found any interpretation of the significance of cycle rank, which may be better converted to other structural parameters that include the minimum path length. [29] Subsequent works toward such directions are ongoing, and the results will be reported elsewhere.

**Acknowledgments**

This study is partly supported by JST-CREST (JPMJCR1992) and JSPS KAKENHI (22H01189).

**References**

1    Takamasa. Sakai, *Physics of Polymer Gels*, Wiley, 2020.
2    C. W. Barney, Z. Ye, I. Sacligil, K. R. McLeod, H. Zhang, G. N. Tew, R. A. Riggleman and A. J. Crosby, *Proceedings of the National Academy of Sciences*, 2022, **119**, 2–7.
3    M. Zhong, R. Wang, K. Kawamoto, B. D. Olsen and J. A. Johnson, *Science (1979)*, 2016, **353**, 1264–1268.
4    J. Lake and A. G. Thomas, *Proc R Soc Lond A Math Phys Sci*, 1967, **300**, 108–119.
5    S. Lin and X. Zhao, *Phys Rev E*, 2020, **102**, 52503.
6    Y. Akagi, T. Katashima, H. Sakurai, U. Il Chung and T. Sakai, *RSC Adv*, 2013, **3**, 13251–13258.
7    T. Fujiyabu, N. Sakumichi, T. Katashima, C. Liu, K. Mayumi, U. Chung and T. Sakai, *Sci Adv*, 2022, **8**, abk0010_1-abk0010_10.
8    Y. Masubuchi, Y. Doi, T. Ishida, N. Sakumichi, T. Sakai, K. Mayumi and T. Uneyama, *Macromolecules*, 2023, **56**, 2217–2223.
9    Y. Masubuchi, Y. Doi, T. Ishida, N. Sakumichi, T. Sakai, K. Mayumi, K. Satoh and T. Uneyama, *Macromolecules*, 2023, **56**, 9359–9367.
10   Y. Masubuchi, *Polym J*, 2024, **56**, 163–171.
11   Y. Masubuchi, *Polymer (Guildf)*, 2024, **297**, 126880.
12   T. Sakai, *Nihon Reoroji Gakkaishi*, 2019, **47**, 183–195.
13   M. Shibayama, X. Li and T. Sakai, *Colloid Polym Sci*, 2019, **297**, 1–12.
14   T. Sakai, Y. Akagi, S. Kondo and U. Chung, *Soft Matter*, 2014, **10**, 6658–6665.

15  Y. Masubuchi, *Nihon Reoroji Gakkaishi*, 2024, **52**, 21–26.

16  R. L. Honeycutt, *Phys Rev A (Coll Park)*, 1992, **45**, 600–603.

17  Y. Masubuchi, R. Yamazaki, Y. Doi, T. Uneyama, N. Sakumichi and T. Sakai, *Soft Matter*, 2022, **18**, 4715–4724.

18  Y. Masubuchi and T. Uneyama, *Soft Matter*, 2019, **15**, 5109–5115.

19  M. Lang, *Macromolecules*, 2019, **52**, 6266–6273.

20  K. Nishi, H. Noguchi, T. Sakai and M. Shibayama, *J Chem Phys*, 2015, **143**, 184905.

21  K. Nishi, M. Chijiishi, Y. Katsumoto, T. Nakao, K. Fujii, U. Chung, H. Noguchi, T. Sakai and M. Shibayama, *J Chem Phys*, 2012, **137**, 224903.

22  J. Lei, Z. Li, S. Xu and Z. Liu, *J Mech Phys Solids*, 2021, **156**, 104599.

23  J. Nocedal, *Math Comput*, 1980, **35**, 773.

24  A. Arora, T. S. Lin and B. D. Olsen, *Macromolecules*, 2022, **55**, 4–14.

25  S. Ishikawa, Y. Iwanaga, T. Uneyama, X. Li, H. Hojo, I. Fujinaga, T. Katashima, T. Saito, Y. Okada, U. Chung, N. Sakumichi and T. Sakai, *Nat Mater*, , DOI:10.1038/s41563-023-01712-z.

26  C. W. Macosko and D. R. Miller, *Macromolecules*, 1976, **9**, 199–206.

27  C. W. Macosko and D. R. Miller, *Makromol. Chem.*, 1991, **192**, 377–404.

28  H. M. James and E. Guth, *J Chem Phys*, 1943, **11**, 455–481.

29  Z. Yu and N. E. Jackson, , DOI:10.48550/arXiv.2405.03551.